# Spectroscopic Evidence for Strong Quantum Spin Fluctuations with Itinerant Character in YFe$_2$Ge$_2$

N. Sirica[1], F. Bondino[2], S. Nappini[2], I. Píš[2,3], L. Poudel[1,4], A. D. Christianson[1,4], D. Mandrus[5,6], D. J. Singh[6] and N. Mannella[1]*

[1]Department of Physics and Astronomy, University of Tennessee, Knoxville, TN 37996, USA
[2]IOM-CNR, S.S. 14 Km 163.5, I-34149 Basovizza (TS), Italy
[3]Elettra - Sincrotrone Trieste S.C.p.A., S.S. 14 Km 163.5, I-34149 Basovizza (TS), Italy
[4]Quantum Condensed Matter Division, Oak Ridge National Laboratory, Oak Ridge, TN 37831, USA
[5] Department of Materials Science and Engineering, University of Tennessee, Knoxville, TN 37996, USA
[6]Materials Science and Technology Division, Oak Ridge National Laboratory, Oak Ridge, TN 37831, USA.
*E-mail:  nmannell@utk.edu

## ABSTRACT

We report x-ray absorption and photoemission spectroscopy of the electronic structure in the normal state of metallic YFe$_2$Ge$_2$.  The data reveal evidence for large fluctuating spin moments on the Fe sites, as indicated by exchange multiplets appearing in the Fe 3s core level photoemission spectra, even though the compound does not show magnetic order.   The magnitude of the multiplet splitting is comparable to that observed in the normal state of the Fe-pnictide superconductors.  This shows a connection between YFe$_2$Ge$_2$ and the Fe-based superconductors even though it contains neither pnictogens nor chalcogens.  The implication is that the chemical range of compounds showing at least one of the characteristic magnetic signatures of the Fe-based superconductors is broader than previously thought.

The interplay between superconductivity and magnetism is one of the most interesting topics in condensed matter physics.  Conventional, i.e. s-wave electron-phonon mediated, superconductivity is damaged by nearness to magnetism [1]. On the other hand, unconventional forms of superconductivity can be realized in proximity to magnetically ordered states, as exemplified by high-T$_C$ cuprates, heavy fermions, and iron-based high temperature superconductors (Fe-HTSC) [2,3].  In the case of Fe-HTSC, there is evidence for a close association between magnetism and superconductivity, and in particular an unconventional spin-fluctuation mediated superconducting state is considered likely [4,5].  Nematicity and its relationship to magnetism have also been discussed as a common feature of Fe-HTSC [6,7].   Two signatures are particularly indicative of the unusual magnetism found in the pnictide Fe-HTSC.  One consists of large discrepancies between the experimental phase diagrams and mean-field-like approximate density functional theory (DFT) calculations, which overestimate the magnetic tendencies [8]. In particular, the antiferromagnetic ordered spin moments were found to be considerably smaller than the DFT predictions of magnetic ground states with large spin moments ($\approx 2\mu_B$) that are largely independent of doping.   The other signature is the presence in the normal non-magnetically ordered states of large fluctuating spin moments detected with fast measurements occurring on sub-picosecond time scales, such as inelastic neutron scattering (INS), x-ray emission (XES) and core-level photoemission spectroscopy (PES) [9,10,11,12,13].  It has been discussed how the interplay of spin and nematic fluctuations in the normal state may be central to high temperature superconductivity [14,15,16].

In this Letter, we report x-ray absorption and photoemission spectroscopy of the electronic structure in the normal state of $YFe_2Ge_2$. The data reveal a striking resemblance with the electronic structure of Fe-pnictide superconductors, including the occurrence of fluctuating spin moments on the Fe sites, as indicated by exchange multiplets appearing in the Fe 3s core level photoemission spectra. The magnitude of the multiplet splitting is similar to that found previously in the normal state of the Fe-pnictide superconductor $CeFeAsO_{0.89}F_{0.11}$ [9]. These findings imply that magnetic behavior similar to the Fe-HTSC can be found in compounds containing neither pnictogens nor chalcogens, and suggest that perhaps unconventional superconductivity related to that in the Fe-pnictides could be found in Fe-Ge compounds.

The $YFe_2Ge_2$ compound is an Fe-containing compound with evidence for nearness to a quantum critical point [17]. $YFe_2Ge_2$ crystallizes in the $ThCr_2Si_2$ type structure (*I4/mmm*), the same structure as the 122 Fe-pnictide Fe-HTSC. In addition, there is evidence for possibly bulk superconductivity with critical temperature $T_C$ = 1.8 K [17]. Based on electron counting, $YFe_2Ge_2$ is electronically akin to $KFe_2As_2$, an Fe-based superconductor with a highly enhanced specific heat, similar to $YFe_2Ge_2$ [18]. The related compound $LuFe_2Ge_2$, which is isoelectronic and has the same structure as $YFe_2Ge_2$, exhibits antiferromagnetic order below 9 K [19]. This magnetic order is continuously suppressed in $Lu_{1-x}Y_xFe_2Ge_2$ as the Y content is increased, with the quantum critical point occurring for $x \approx 0.2$ [20]. The proximity of the end series compound $YFe_2Ge_2$ to quantum criticality is consistent with the non-Fermi liquid behavior of the specific heat and resistivity [17].

DFT calculations found magnetic ground states [21,22], in contrast to the fact that $YFe_2Ge_2$ is a metal that does not exhibit magnetic order. The DFT calculations show competition between different antiferromagnetic states, including an A-type antiferromagnetic structure consisting of ferromagnetic Fe-planes with antiferromagnetic stacking along the *c*-axis and, as shown by Subedi [21], a stripe-like structure similar to the Fe-based superconductors. The overestimation of the magnetic tendencies within DFT is unusual, since in general DFT underestimates the magnitude of the ordered spin moment in correlated materials and normally gives reliable predictions in the absence of strong correlations. It is not, however, unprecedented: It suggests proximity to, or incipient magnetism. In particular, besides the pnictide Fe-HTSC [3,8], it occurs in several compounds near quantum critical points associated with itinerant magnetism [23], as might be anticipated based on theoretical arguments [24,25,26]. In strong analogy to the pnictide Fe-HTSC, our x-ray spectroscopy data reveal in the normal state of $YFe_2Ge_2$ the presence of both itinerant electrons and large fluctuating spin moments on the Fe sites.

High quality crystalline samples were grown out of Sn flux. X-ray powder diffraction patterns were found to be consistent with that of the tetragonal crystal structure with space group *I4/mmm*. The PES and XAS measurements were carried out on the BACH beamline at the Elettra Synchrotron Facility. Several samples have been measured at room temperature in a pressure better than $8 \times 10^{-10}$ mbar. Surface cleanliness was assured by periodic (every 20 minutes) *in-situ* scraping with a diamond file [27]. Quantitative PES analysis of core-level spectra showed no surface contamination during spectra acquisition.

The Fe 2p core level PES and the Fe $L_{23}$ XAS spectra, shown in Fig.1, are remarkably similar to those excited in some of the pnictide Fe-HTSC such as $CeFeAsO_{0.89}F_{0.11}$ [9] and $Ba(Fe_{1-x}Co_x)_2As_2$ [28]. As in the pnictide Fe-HTSC, the Fe 2p PES and Fe $L_{23}$ XAS spectra display signatures which are typical of delocalized, itinerant electrons, in agreement with other studies [29,30]. Specifically, the Fe 2p PES spectrum does not show the additional satellite structures indicative of the presence of strong electron correlation and localization effects, as for example found in the cuprates. Rather, the PES Fe 2*p* is akin to that of Fe metal and intermetallic compounds [31]. The Fe $L_{23}$ XAS spectrum [Fig. 1 (b)] does not show the presence of well-defined multiplet structures. The broad and weak shoulder at $\approx$ 705 eV, also found in the Fe-HTSC, is most likely indicative of the covalent nature of the Ge and Fe conduction band states. Indeed, such structure is present in the Fe XAS spectra of Fe–X (X is an *sp*-element) compounds with strong Fe 3*d*-X n*p* hybridization such as Fe silicides [32].

case for the Ge-4p and Ge-4s states, due to their extreme delocalization. This preempts the extraction of the precise Ge-4s and Ge-4p contribution to the DOS, which is prerequisite for including the cross-sections.

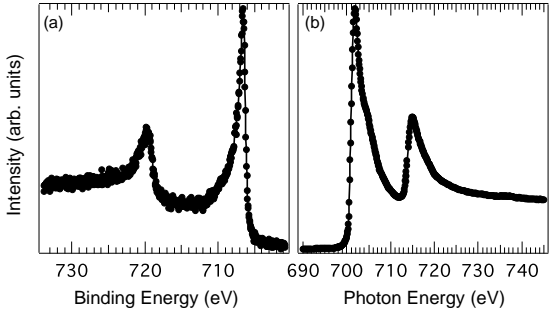

**Fig. 1.** (a) Fe 2*p* core level PES spectrum excited with photon energy hν = 907.5 eV. (b) Fe $L_{23}$ XAS spectrum. Note the marked similarities with corresponding spectra excited in Fe-based high temperature superconductors, as reported in ref. [9].

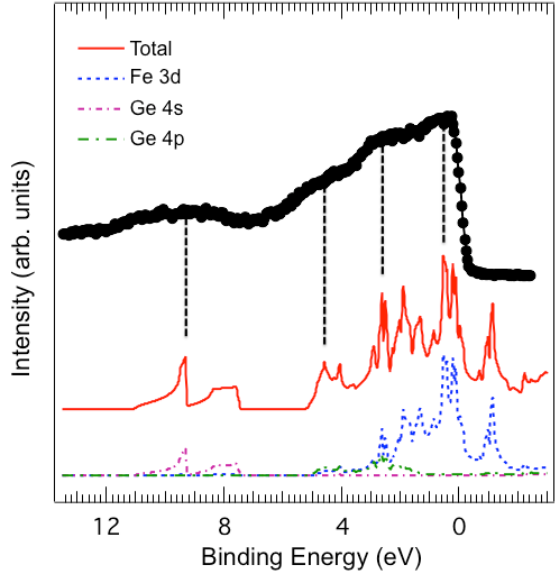

**Fig. 2.** VB measured with photon energy hν = 907.5 eV. Note the high DOS at $E_F$, denoted with the vertical dashed line. Also shown are the calculated total DOS, and the Fe 3d, Ge 4s and Ge 4p partial DOS.

Fig. 2 (a) shows the valence band (VB) PES spectrum of YFe$_2$Ge$_2$. Given that the resulting samples are polycrystalline due to in-situ scraping of the surface, this spectrum provides a representation of the occupied density of states (DOS), modulated by atomic cross-section effects and instrumental resolution (≈ 500 meV). Similar to the Fe-HTSC, there is a high DOS at the Fermi level ($E_F$) primarily of Fe-derived character. This is consistent with the lineshape of the Fe 2*p* PES spectrum: The high Fe DOS at $E_F$ is very effective in completely screening the Fe 2*p* core-hole excitation, leading to an absence in the core-level spectra of satellite structures. Overall, there is a very good correspondence of the main features in the VB with the total DOS and its decomposition into its main three components, i.e. Fe-3d, Ge-4p and Ge-4s, as calculated in ref. [22]. A more detailed comparison requires including the proper orbital dependent cross-sections for the Fe-3d, Ge-4p and Ge-4s states. This task is complicated by the fact that the orbital decomposition in the LAPW method relies on projections onto the LAPW spheres. The Fe 3d states are almost entirely contained in the LAPW sphere, making this a good approximation. On the contrary, this is not the

Perhaps the similarities of the PES and XAS spectra shown in Figs. 1 and 2 should not be that surprising, in light of the fact that both YFe$_2$Ge$_2$ and the 122 Fe-HTSC have the same crystal structure. On the other hand, the 122 Fe-HTSC are pnictides, i.e. they contain pnictogen elements, and do not contain Ge. The similarity with Fe-based compound containing Ge is therefore not necessarily expected.

The PES and XAS spectra shown in Figs. 1 and 2 are not the only indication of a marked similarity between YFe$_2$Ge$_2$ and the pnictide Fe-HTSC. As in the pnictides, the Fe 3*s* core-level spectrum in YFe$_2$Ge$_2$ exhibits a multiplet splitting (M-SP) of the binding energy (BE). M-SP effects arise from the exchange coupling of the core 3*s* electron with the net spin $S_V$ in the unfilled 3*d*/4*s* shells of the emitter atom, Fe in this case. Since M-SP occurs exclusively in atoms with the outer subshell(s) partially occupied with a non-vanishing net spin $S_V$, the Fe 3s spectrum in Fig. 3(a) indicates the presence of spin moments on the Fe sites [9,12,13]. The multiplet energy separation $\Delta E_{3s}$ permits estimation of the effective net spin of the emitter atom, i.e., the local spin moment. It has been shown that for itinerant systems the net spin $S_V$ can be found by extrapolating the linear fit of the measured splitting $\Delta E_{3s}$ for ionic compounds versus $(2S_V + 1)$ (33,34). This approach has been used for itinerant magnetic systems, including the pnictide Fe-HTSC [9,12]. Using this same approach, the value of the measured splitting $\Delta E_{3s} = 3.15$ eV obtained with a two-component fit of the Fe 3s spectrum (cf. Fig. 3(a)) provides a value of the Fe spin moment $2S_V \approx 1$ $\mu_B$ [35]. On the contrary, the Ge 4s spectrum does not seem to show any splitting, given the reasonably good fit obtained with a single Voigt function (cf. Fig. 3(b)). A two-peaks fit would obviously give a better chi-squared, but the splitting turns out to be very small, indicating that, if any, the spin polarization of Ge is very small.

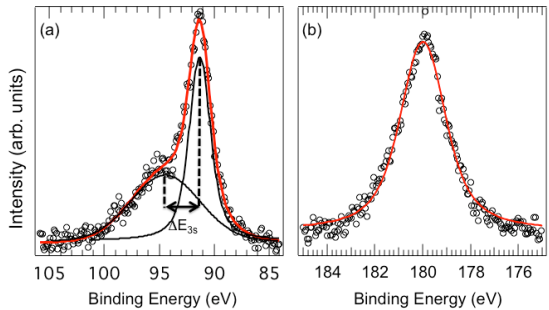

**Fig. 3.** (a) Fe 3s and (b) Ge 3s core level PES spectra excited with photon energy hv = 907.5 eV. The value of the energy separation of the multiplet is $\Delta E_{3s} = 3.15$ eV

The extremely short time scales involved in the photoemission process ($10^{-17}$ - $10^{-16}$ s) can account for the disagreement with conventional static magnetic measurements, according to which Fe in YFe$_2$Ge$_2$ is non magnetic, with a Pauli paramagnetic susceptibility [17]. This indicates that, as for the pnictide Fe-HTSC, the value $\approx 1$ $\mu_B$ extracted with PES is an estimate of the averaged magnitude of fast-fluctuating spin moments on the Fe sites. As for the Fe-HTSC, the best fits to the Fe 3s spectra are always obtained when the curve fitting the peak at higher BE is mainly of Gaussian character, with a width much larger than that of the lower BE peak and that expected from experimental resolution. Indeed, a system with strong itinerant Fe-based spin fluctuations would locally mimic fluctuations in the magnitude of the moment on Fe sites, which should appear in an Fe 3s spectrum as sidebands at higher *BE* with the envelope of the peaks being a Gaussian, reflecting the normal character of their distribution. This is remarkably different from spin-fluctuations associated with paramagnetism due to local moments of fixed magnitude as found in most Fe compounds, since in this case all of the Fe local spin moments would exhibit the same constant value, with two peaks of similar width. Signatures of electron itinerancy and fluctuating Fe spin moments similar to those found here for YFe$_2$Ge$_2$ are also found in pnictide Fe-HTSC and other Fe-intermetallic compounds that are either near quantum critical points (FeAl [36], NbFe$_2$ [37]), and/or were subsequently found to have ordered magnetic ground states (NbFe$_2$,TiFe$_2$). [38,39,40].

One characteristic that may be expected for magnetism with itinerant character, as suggested for the Fe-based superconductors [41], is the presence of longitudinal spin fluctuations in the ordered state. We extended our previous calculations reported in Ref. [22], to include a stripe-like structure similar to the Fe-HTSC [21] and an additional order with the same in-plane arrangement, but antiferromagnetic stacking along the *c*-axis. The approach is the same as that used previously, i.e. a PBE generalized gradient approximation. Energies per formula unit relative to the non-spin-polarized state were calculated for ferromagnetic (-120 meV), A-type antiferromagnetic (-137 meV), C-type (-26 meV, (see ref. [22]), G-type (-4 meV), Fe-pnictide-like stripes with ferromagnetic stacking (-129 meV) and Fe-pnictide-like stripes stacked antiferromagnetically along *c* (-161 meV). These indicate, following the arguments of Ref. [21] and [22], that the magnetism has itinerant character. Additionally, there is a clear competition between the A-type and stripe-like states. Although the antiferromagnetic stacked stripe has the lowest energy, its origin in nesting of relatively small pockets as discussed by Subedi [21] means that it is a sharper feature in momentum space. As such, it may be more strongly influenced by scattering and more readily suppressed by spin-fluctuations. This is similar to a proposed scenario in the triplet superconductor $Sr_2RuO_4$ [42,43,44] where competition between different magnetic orders [45] is important for suppressing magnetism in favor of an unconventional superconducting state. Both the lowest energy $c$-axis stacked stripe order and the A-type antiferromagnetic orders couple strongly to electrons at the Fermi energy, as indicated by the DOS $N(E_F)$, which is reduced by magnetic ordering to 51% and 68% of the non-spin polarized value, respectively.

The normal state of $YFe_2Ge_2$ is markedly similar to that of pnictide Fe-HTSC. It appears to be quite unique: It features i) no signatures of strong local Mott-Hubbard type correlations analogous to the cuprate HTSC (cf. Fig. 1), ii) an itinerant Fe d-band character (cf. Fig. 2), iii) a high DOS at $E_F$ (cf. Fig. 2), iv) an overestimation of the magnetic tendencies within DFT, and (v) the occurrence of quantum fluctuations of the Fe spin moment as revealed by multiplet splitting of the binding energy of Fe 3s core level PES spectra (cf. Fig. 3). The 1 $\mu_B$ value of the Fe spin is similar to that found previously in the normal state of the pnictide Fe-HTSC $CeFeAsO_{0.89}F_{0.11}$ [9,12]. In the pnictide Fe-HTSC, it has been proposed that the occurrence of quantum fluctuations of the spin moment could be responsible for the DFT overestimation of the ordered spin moments [46,47,48]. There is growing evidence that strong fluctuations of the spin moment are an important defining characteristic of the normal state of the pnictides. Recent INS, XES, PES data have provided evidence of fluctuating Fe spin moments in both ordered and paramagnetic phases of several pnictide compounds [10,11,12,13]. The values of the Fe spin moments are much larger than those previously found with conventional magnetic measurements such as nuclear magnetic resonance (NMR), muon spin resonance (μ-SR) and Mössbauer spectroscopy [3,13]. The detection of fluctuating spin moments was possible because INS, XES and PES measurements occur on sub-picosecond time scales, much faster than the $10^{-8}$ s - $10^{-6}$ s time scales typical of NMR, $\mu$–R and Mössbauer.

In light of these facts, the apparent similarity of the magnetic (fluctuating moments), crystallographic, chemical, and electronic degrees of freedom of $YFe_2Ge_2$ to the Fe-HTSC is important. Although there are some important differences with respect to the pnictides, such as a nearby magnetic order of different nature and an unusually high specific heat, $YFe_2Ge_2$ is a compound that shares many of the characteristics found in the parent compounds of pnictides that were later found to host high temperature superconductivity. Based on this, one might speculate that high temperature superconductivity could be found in compounds based on Fe and Ge. Certainly, it will be of considerable interest to better elucidate the magnetic behavior of metallic Fe-germanide compounds and their similarities and differences from the Fe-based superconductors.

This work was supported by the National Science Foundation, Division of Material Research, grant DMR-1151687 (N.M.). A.D.C. and L.P. are supported by the scientific User Facilities Division, Office of Basic Energy Sciences, U.S. Department of Energy (DOE).